\RequirePackage{ifpdf}
\ifpdf 
\documentclass[pdftex]{sigma}
\else
\documentclass{sigma}
\fi

\begin{document}

\renewcommand{\labelenumi}{(\alph{enumi})}

\allowdisplaybreaks

\renewcommand{\thefootnote}{$\star$}

\renewcommand{\PaperNumber}{103}

\FirstPageHeading

\ShortArticleName{Equivalent and Alternative Forms for BF Gravity with Immirzi Parameter}

\ArticleName{Equivalent and Alternative Forms for BF Gravity \\ with Immirzi Parameter\footnote{This
paper is a contribution to the Special Issue ``Loop Quantum Gravity and Cosmology''. The full collection is available at \href{http://www.emis.de/journals/SIGMA/LQGC.html}{http://www.emis.de/journals/SIGMA/LQGC.html}}}

\Author{Merced MONTESINOS and Mercedes VEL\'AZQUEZ}

\AuthorNameForHeading{M.~Montesinos and M.~Vel\'azquez}

\Address{Departamento de F\'{\i}sica, Cinvestav, Instituto Polit\'ecnico Nacional 2508,\\
San Pedro Zacatenco, 07360, Gustavo A. Madero, Ciudad de M\'exico, M\'exico}
\Email{\href{mailto:merced@fis.cinvestav.mx}{merced@fis.cinvestav.mx}, \href{mailto:mquesada@fis.cinvestav.mx}{mquesada@fis.cinvestav.mx}}
\URLaddress{\url{http://www.fis.cinvestav.mx/~merced/}}

\ArticleDates{Received August 31, 2011, in f\/inal form November 07, 2011;  Published online November 11, 2011}

\Abstract{A detailed analysis of the BF formulation for general relativity given by Capo\-villa, Montesinos, Prieto, and
Rojas is performed. The action principle of this formulation is written in an equivalent form by doing a transformation
of the f\/ields of which the action depends functionally on. The transformed action principle involves two BF terms and
the two Lorentz invariants that appear in the original action principle generically. As an application of this
formalism, the action principle used by Engle, Pereira, and Rovelli in their spin foam model for gravity is recovered
and the coupling of the cosmological constant in such a formulation is obtained.}


\Classification{83C05; 83C45}

\section{Introduction}

{\sloppy One of the main challenges nowadays is to establish links between loop quantum gravity (LQG)~\cite{lqg} and spin foam
models~\cite{spinfoam}, which are the main approaches to the nonperturbative and background-independent quantizations
of general relativity. Whether or not the two quantization schemes yield two dif\/ferent quantum theories is still an
open problem (see~\cite{spin-loops}). The search for the links between these approaches lies mostly in the quantum
realm, but there are still some aspects of this correspondence that are unclear classically. It is possible to say that
the connection between the two frameworks at the classical level is the relationship between the Holst's action and the
BF formulations for general relativity because LQG is based at the classical level on Holst's action but spin foam
models for gravity are related to constrained BF theories. This is the issue studied in this paper.

}

General relativity expressed as a constrained BF theory was given by Pleba\'nski many years ago~\cite{pleb77}. The
basic idea behind the Pleba\'nski formulation is that the fundamental variables for describing the gravitational f\/ield
(general relativity) are neither a metric (as it is in the Einstein--Hilbert action) nor a tetrad together with a
Lorentz connection (as it is in the Palatini action), but rather two-form f\/ields, a connection one-form, and some
Lagrange multipliers. The geometry of spacetime is built up from these fundamental blocks. In order to bring tetrads
into the formalism, the two-forms are eliminated by solving an equation among them, which implies that the two-forms
can be expressed in terms of tetrad f\/ields, and by inserting back this expression for the two-forms into the
Pleba\'nski action, it becomes the self-dual action for general relativity~\cite{1Samuel,1Jacobson}.

This view point has been adopted in the construction of other action principles, which also express general relativity
as a constrained BF theory~\cite{penrose,capo,rob,cqg1999,cqg1999b,cqgl2001}. The link with tetrad gravity is again
made by solving the constraints for the two-form f\/ields which amounts to express them in terms of tetrad f\/ields. In
particular, a formulation for real general relativity expressed as a constrained BF theory that involves the Immirzi
parameter~\cite{barbero,immirzi,Holst} was given in~\cite{cqgl2001} by Capovilla, Montesinos, Prieto, and Rojas
(hereafter CMPR formulation). It is well-known that the Immirzi parameter in such a formulation appears naturally when
the two Lorentz invariants $\eta_{IK}\eta_{JL}-\eta_{IL}\eta_{JK}$ and $\varepsilon_{IJKL}$ are introduced in the
constraint on the Lagrange multipliers $\phi_{IJKL}$. This action principle involves just {\it one} BF term. In this
work it is shown that by performing a suitable transformation on the f\/ields involved in the theory, the original action
principle can be written in a form that involves {\it two} BF terms, one of them containing a parameter that will be
identif\/ied {\it a posteriori} with the Immirzi parameter (see also~\cite{oriti}). This allows us to relate the
CMPR formulation with dif\/ferent real BF formulations of gravity currently employed in the literature (see e.g.~\cite{epr}). Furthermore, the same transformation is applied to the action principle studied in~\cite{prd2010},
which includes the cosmological constant, and we obtain the coupling of the cosmological constant in the framework of~\cite{epr}. The material reported in this paper is part of the work presented in~\cite{mpvq2}.

\section{CMPR action for gravity}
The action principle for pure gravity introduced by Capovilla, Montesinos, Prieto, and Rojas in~\cite{cqgl2001} is
given by
\begin{gather}
S [Q,A,\psi,\mu] =\int_{\mathcal{M}^4} \Big[ Q^{IJ} \wedge F_{IJ} [A] - \frac12 \psi_{IJKL} Q^{IJ} \wedge Q^{KL}
 \nonumber \\
 \hphantom{S [Q,A,\psi,\mu] =\int_{\mathcal{M}^4}\Big[}{} -
\mu  \left ( a_1  \psi_{IJ}{}^{IJ} + a_2  \psi_{IJKL}  \varepsilon^{IJKL} \right )
\Big],\label{cmpr1}
\end{gather}
where $A^I{}_J$ is an Euclidean or Lorentz connection one-form, depending on whether $SO(4)$ or $SO(3,1)$ is taken as
the internal gauge group, and $F^I{}_J[A]= d A^I{}_J + A^I{}_K \wedge A^K{}_J$ is its curvature; the $Q$'s are a set of
six two-forms on account of their antisymmetry $Q^{IJ}=-Q^{JI}$; the Lag\-range multiplier $\psi_{IJKL}$ has 21
independent components due to the properties $\psi_{IJKL}=\psi_{KLIJ}$, $\psi_{IJKL}=-\psi_{JIKL}$, and
$\psi_{IJKL}=-\psi_{IJLK}$; the Lagrange multiplier $\mu$ implies the additional restriction $a_1  \psi_{IJ}{}^{IJ} +
a_2  \psi_{IJKL}  \varepsilon^{IJKL}=0$ on the Lagrange multiplier $\psi_{IJKL}$. The Lorentz (Euclidean) indices
$I,J,K,\ldots=0,1,2,3$ are raised and lowered with the Minkowski (Euclidean) metric $(\eta_{IJ})=
\mbox{diag}(\sigma,+1,+1,+1)$ where $\sigma=+1$ for Euclidean and $\sigma=-1$ for Lorentzian signatures, respectively.

The variation of the action (\ref{cmpr1}) with respect to the independent f\/ields gives the equations of motion
\begin{gather}
\delta Q: \ F_{IJ} [A] - \psi_{IJKL} Q^{KL}=0, \nonumber \\
\delta A: \ D Q^{IJ} =0, \nonumber\\
\delta \psi: \ Q^{IJ} \wedge Q^{KL} + 2 a_1 \mu  \eta^{[I\mid K \mid} \eta^{J]L} + 2 a_2 \mu  \varepsilon^{IJKL}=0,\label{delta psi}\\
\delta \mu: \ a_1  \psi_{IJ}{}^{IJ} + a_2  \psi_{IJKL} \varepsilon^{IJKL} =0.\nonumber
\end{gather}
By contracting equation (\ref{delta psi}) with the Killing--Cartan metric $\eta_{[I\mid K \mid}
\eta_{J]L}=\frac12(\eta_{IK}\eta_{JL}-\eta_{IL}\eta_{JK})$ and $\varepsilon_{IJKL}$, one gets $a_1 \mu= - \frac{1}{12}
Q^{IJ} \wedge Q_{IJ}$ and $a_2 \mu = - \frac{\sigma}{4!} Q^{IJ} \wedge {^{\ast}Q}_{IJ}$ respectively, where
${^*}Q^{IJ}:=\frac12 \varepsilon^{IJ}{}_{KL} Q^{KL}$. The non-degenerate case corresponds to $\mu \neq 0$ whereas the
degenerate case corresponds to $\mu = 0$. Let us restrict the analysis to the non-degenerate case. Inserting back $a_1
\mu$ and $a_2 \mu$ into (\ref{delta psi}), it is obtained
\[
Q^{IJ} \wedge Q^{KL} - \frac{1}{6} \left ( Q^{MN} \wedge
Q_{MN} \right) \eta^{[I\mid K \mid} \eta^{J]L} - \frac{2\sigma}{4!} \left ( Q^{MN} \wedge {^{\ast}Q}_{MN} \right )
\varepsilon^{IJKL}=0,
\]
together with
\[
2 a_2 Q^{IJ} \wedge Q_{IJ} - \sigma a_1 Q^{IJ} \wedge {^{\ast}Q}_{IJ}=0,
\]
that follows from the equality of the two expressions for $\mu$ and the fact that $a_1 \neq 0$ and $a_2 \neq 0$. It is
shown in~\cite{cqgl2001} that
\begin{gather}\label{sol}
Q^{IJ}=\alpha \, {}^{\ast} \big(e^I \wedge e^J\big) + \beta  e^I \wedge e^J,
\end{gather}
is the general solution for the $Q$'s provided that the constants $\alpha$ and $\beta$ satisfy
\begin{gather}\label{quotien}
\frac{a_2}{a_1}  =  \frac{\alpha^2 + \sigma \beta^2}{4 \alpha\beta}.
\end{gather}
By inserting the solution (\ref{sol}) into the action principle (\ref{cmpr1}), we get
\[
S[e,A]=\int_{\mathcal{M}^4} \left [ {^{\ast}} \left ( e^I \wedge e^J \right ) +  \frac{\beta}{\alpha}  e^I \wedge e^J
\right ] \wedge F_{IJ} [A].
\]
Notice that, as remarked in~\cite{cqgl2001}, the Immirzi parameter appears naturally in equation~(\ref{sol})
because the two invariants $\psi_{IJ}{}^{IJ}$ and $\psi_{IJKL}\varepsilon^{IJKL}$ are present in the action.

\subsection[CMPR formulation with $a_1=0$ or $a_2=0$]{CMPR formulation with $\boldsymbol{a_1=0}$ or $\boldsymbol{a_2=0}$}
The cases when $a_1$ or $a_2$ are equal to zero have been analyzed separately \cite{capo,cqg1999,cqg1999b}. In
particular, if $a_1=0$ and $a_2\neq0$ the Lorentz invariant $\psi_{IJ}{}^{IJ}$ is not present in the action, which
reduces to
\begin{gather}\label{epsilon}
S[Q,A,\psi,\mu] =\int_{\mathcal{M}^4} \left [ Q^{IJ} \wedge F_{IJ} [A] - \frac12 \psi_{IJKL} Q^{IJ} \wedge Q^{KL}
-\mu  \psi_{IJKL}\varepsilon^{IJKL} \right ].
\end{gather}
After solving the constraint on the $Q$'s, they can be written in terms of the tetrad $e^I$ as
\begin{gather}\label{pietriEXTRA}
(i) \quad Q^{IJ}=\kappa_1   {^{\ast}} \left ( e^I \wedge e^J \right ), \qquad \quad (ii) \quad Q^{IJ}=\kappa _2  e^I
\wedge e^J,
\end{gather}
where $\kappa_1$, $\kappa_2$ are constants. By inserting these expressions for the $Q$'s into (\ref{epsilon}), we get
action principles for two dif\/ferent theories, one of which is general relativity~\cite{cqg1999,cqg1999b}
\begin{gather}
 (i) \quad S_{1} [e,A] = \kappa_1 \int_{\mathcal{M}^4}  {^{\ast}} \left( e^I \wedge e^J \right) \wedge F_{IJ} [A],
 \nonumber\\ \label{epsilonENes}
 (ii) \quad S_{2} [e,A] = \kappa_2 \int_{\mathcal{M}^4} \left ( e^I \wedge e^J \right ) \wedge F_{IJ} [A].
\end{gather}
In~\cite{vlad} was proposed to consider the action $S_2[e,A]$ as a genuine f\/ield theory in its own right because
it is background-independent and dif\/feomorphism-invariant. Recently, it has been shown that $S_2[e,A]$ is indeed
topological if the spacetime $\mathcal{M}^4$ has no boundary. The proof is given by performing the covariant canonical
analysis to this action as well as by doing the Dirac's canonical analysis with and without breaking local Lorentz
invariance~\cite{liu,maga,liuthesis}. The relevance of the action $S_2[e,A]$ is not academic, this f\/ield theory is a
topological limit of general relativity obtained by taking the Newton constant $G \rightarrow \infty$ and the Immirzi
parameter $\gamma \rightarrow 0$ while keeping the product $G \gamma$ constant~\cite{liu}.

On the other hand, if $a_1\neq0$ and $a_2=0$, the action takes the form \cite{capo}
\begin{gather}\label{traza}
S[Q,A,\psi,\mu] =\int_{\mathcal{M}^4} \left [ Q^{IJ} \wedge F_{IJ} [A] - \frac12 \psi_{IJKL} Q^{IJ} \wedge Q^{KL}
-\mu  \psi_{IJ}{}^{IJ} \right ].
\end{gather}
The solutions for the two-forms in terms of the tetrad $e^I$ are given by
\[
Q^{IJ}=\kappa \left[ {^{\ast}} \left ( e^I \wedge e^J \right )\pm \sqrt{-\sigma}  e^I \wedge e^J\right].
\]
Therefore, the action principle in terms of the tetrad takes the form
\[
S_E[e,A] = \kappa \int_{\mathcal{M}^4}  \left [ {^{\ast}} \left ( e^I \wedge e^J \right ) \pm i e^I \wedge e^J \right ]
\wedge F_{IJ} [A]
\]
in the Euclidean case whereas in the Lorentzian case it becomes
\begin{gather*}
S_{L}[e,A]= \kappa \int_{\mathcal{M}^4} \left[{^{\ast}} \left ( e^I \wedge e^J \right ) \pm e^I \wedge e^J \right]
\wedge F_{IJ} [A].
\end{gather*}
Notice that last form of the action includes the coupling of the term $e^I \wedge e^J \wedge F_{IJ}[A]$  added by Holst
to the Palatini action with (what it is called now) Immirzi parameter equal to $\pm 1$, but this form of the action was
reported in~\cite{capo} several years before Holst, Immirzi, and Barbero's papers.

\section{Transformation of the CMPR action for gravity} \label{trasfCMPR}

The goal of this section is to study the CMPR formulation for general relativity by performing a linear transformation
from the original variables the action principle depends functionally on to a new set of two-forms and Lagrange
multipliers. It will be shown that the resulting action principle involves the {\it two} possible BF terms, $B^{IJ}
\wedge F_{IJ} [A]$ and ${^{\ast}B}^{IJ} \wedge F_{IJ}[A]$ that can be built when the internal gauge group is $SO(3,1)$
or $SO(4)$, with the corresponding change in the symplectic structure as it was pointed out in~\cite{iopmm} and~\cite{mmv}.

In order to do what we have explained, i.e., the alternative writing of the CMPR action, we def\/ine
\begin{gather}\label{transf inversa}
Q^{IJ} := b_1 B^{IJ} + b_2 \,^{\ast} B^{IJ},
\end{gather}
with $b_1$ and $b_2$ constants, from which it follows the inverse transformation
\begin{gather}\label{transformacion}
B^{IJ} = \frac{1}{b_1^2-\sigma b_2^2}  \big(b_1 Q^{IJ} - b_2  {^{\ast} Q}^{IJ} \big),
\end{gather}
provided that
\begin{gather}\label{condition}
b_1^2 -\sigma b_2^2 \neq 0,
\end{gather}
holds.

Using (\ref{transf inversa}), the Lagrangian of action principle (\ref{cmpr1}) acquires the form
\begin{gather}
\big(b_1 B^{IJ} + b_2\, {^*}B^{IJ}\big) \wedge F_{IJ} [A] - \frac12  \psi_{IJKL} \big( b_1 B^{IJ} + b_2\,{^{ \ast}
B}^{IJ} \big)  \wedge \big( b_1 B^{KL} - b_2\, {^{\ast} B}^{KL} \big)
 \nonumber \\  \label{sust parcial}
\qquad- \mu  \big(a_1 \psi_{IJ}{}^{IJ} + a_2  \psi_{IJKL} \varepsilon^{IJKL}\big) ,
\end{gather}
that can further be rewritten by def\/ining
\begin{gather}\label{phi en psi}
\phi_{IJKL} := b_1^2\psi_{IJKL} +b_1 b_2 {^{\ast}\psi}_{IJKL} +b_1 b_2 \psi^{\ast}{}_{IJKL} + b_2^2 \,{^{\ast}
\psi^{\ast}}{}_{IJKL},
\end{gather}
where ${^{\ast}\psi}_{IJKL} := \frac12 \varepsilon^{MN}{}_{IJ}  \psi_{MNKL}$ and ${\psi^{\ast}}{}_{IJKL} := \frac12
\varepsilon^{MN}{}_{KL}  \psi_{IJMN}$ are the dual on the f\/irst and on the second pair of Lorentz indices,
respectively. From (\ref{phi en psi}) it follows that
\begin{gather}\label{psi en phi}
\psi_{IJKL} = \frac1{(b_1^2-\sigma b_2^2)^2} \left( b_1^2 \phi_{IJKL} -b_1 b_2 \,{^{\ast}\phi}_{IJKL} -b_1 b_2
\phi^{\ast}{}_{IJKL} + b_2^2\, {^{\ast} \phi^{\ast}}{}_{IJKL} \right).
\end{gather}
Therefore, the second term of (\ref{sust parcial}) takes the form
\[
\psi_{IJKL} \big( b_1 B^{IJ} + b_2\,{^{ \ast} B}^{IJ} \big) \wedge \big( b_1 B^{KL} - b_2\, {^{\ast} B}^{KL}
\big)=\phi_{IJKL} B^{IJ} \wedge B^{KL}.
\]
Furthermore, using (\ref{psi en phi}), the two invariants $\psi_{IJ}{}^{IJ}$ and $\psi_{IJKL}\varepsilon^{IJKL}$ can be
written in terms of the two invariants $\phi_{IJ}{}^{IJ}$ and $\varepsilon_{IJKL} \phi^{IJKL}$ as
\begin{gather}\label{psi}
\psi_{IJ}{}^{IJ}=\frac1{\big(b_1^2-\sigma b_2^2\big)^2}\left[\big(b_1^2+\sigma b_2^2\big)\phi_{IJ}{}^{IJ}-b_1 b_2
\phi_{IJKL}\varepsilon^{IJKL}\right],
\end{gather}
and
\begin{gather}\label{psi epsilon}
\psi_{IJKL}\varepsilon^{IJKL}=\frac1{\big(b_1^2-\sigma b_2^2\big)^2}\left[\big(b_1^2+\sigma
b_2^2\big)\phi_{IJKL}\varepsilon^{IJKL}-4\sigma b_1 b_2 \phi_{IJ}{}^{IJ}\right].
\end{gather}
Thus, using (\ref{psi}) and (\ref{psi epsilon}), the last term of (\ref{sust parcial}) acquires the form
\begin{gather}\label{H transformada}
a_1\psi_{IJ}{}^{IJ} + a_2\psi_{IJKL}\varepsilon^{IJKL}= A_1 \phi_{IJ}{}^{IJ}+ A_2\phi_{IJKL}\varepsilon^{IJKL},
\end{gather}
with
\begin{gather}
A_1 = \frac1{\big(b_1^2-\sigma b_2^2\big)^2}\left[ a_1\big(b_1^2+\sigma b_2^2\big)-4\sigma a_2 b_1 b_2\right], \nonumber\\
\label{config1} A_2 = \frac1{\big(b_1^2-\sigma b_2^2\big)^2}\left[a_2\big(b_1^2+\sigma b_2^2\big)-a_1 b_1 b_2\right].
\end{gather}

By using the previous steps, the form that the CMPR action (\ref{cmpr1}) acquires once the transformation def\/ined in
equations (\ref{transf inversa}), (\ref{transformacion}) and in equations (\ref{phi en psi}), (\ref{psi en phi}) has been done is
\begin{gather}
 S[B,A,\phi,\mu] =\int_{\mathcal{M}^4} \Big[ \left(b_1 B^{IJ}+ b_2 {^*}B^{IJ}\right) \wedge F_{IJ} [A]
 - \frac12 \phi_{IJKL} B^{IJ} \wedge B^{KL}  \nonumber \\
\hphantom{S[B,A,\phi,\mu] =\int_{\mathcal{M}^4} \Big[}{}  -  \mu  \left ( A_1  \phi_{IJ}{}^{IJ}
 + A_2  \phi_{IJKL}{} \varepsilon^{IJKL} \right ) \Big]. \label{accion transformada}
\end{gather}
Due to the fact that the transformation is invertible, both actions (\ref{cmpr1}) and (\ref{accion transformada}) are
equivalent.

The action principle (\ref{accion transformada}) can still be written in terms of tetrads and a Lorentz connection by
solving the constraint on the $B$'s coming from it. Alternatively, the expression for the two-forms~$B^{IJ}$ in terms
of the tetrad f\/ield can be obtained from the expression for the $Q$'s given in~(\ref{sol}) and from the use of the equation~(\ref{transformacion}).

The relationship between the action principle (\ref{cmpr1}) and (\ref{accion transformada}) will be analyzed in detail
in an example given in Section~\ref{particular transformation}. Some remarks follow:
\begin{enumerate}\itemsep=0pt
\item
It can be observed from equations (\ref{psi}), (\ref{psi epsilon}), (\ref{H transformada}), and (\ref{config1}) that even
though we had started from action (\ref{cmpr1}) with either $a_1=0$ or $a_2=0$, it might be possible to obtain
generically the two Lorentz invariants in the transformed action (\ref{accion transformada}).
\item \label{item b}
It is possible to get just one of the invariants in the transformed action (\ref{accion transformada}) by imposing
either $A_1=0$ or $A_2=0$. For instance, the case $A_1=0$ can be achieved by solving for the ratio $b_2/b_1$ in terms
of the ratio $a_2/a_1$, i.e.\ by choosing a particular transformation (encoded in $b_1$ and $b_2$) and leaving $a_1$ and
$a_2$ arbitrary. Alternatively, $A_1=0$ can be achieved by solving for the ratio $a_2/a_1$ in terms of the ratio
$b_2/b_1$, i.e.\ by choosing a~particular form for the ratio $a_2/a_1$ and leaving the transformation arbitrary.
Similarly, the case $A_2=0$ can also be handled in two analogous ways.
\end{enumerate}

The previous analysis points out that is not correct to refer to the term ${^{\ast}B}^{IJ} \wedge F_{IJ}$ as ``Holst's
term'' simply because the term added by Holst to the Palatini action and given by $e^I \wedge e^J \wedge F_{IJ}$~\cite{Holst} (see also~\cite{capo}) is at the level of tetrads $e^I$ and {\it not} at the level of BF theories.
Even though they might be related, they are not exactly the same thing. In particular, ${^{\ast}B}^{IJ} \wedge F_{IJ}$
could be proportional to $^{\ast} \left ( e^I \wedge e^J \right ) \wedge F_{IJ}$, or to $e^I \wedge e^J \wedge F_{IJ}$,
or to something else depending on the expression for the $B$'s that solves the constraint among them, i.e., {\it a
priori} there is not guarantee that ${^{\ast}B}^{IJ} \wedge F_{IJ}$ would lead to the term added by Holst, because this
will ultimately depend on the expression for the $B$'s in terms of the tetrads.

\subsection{A particular transformation}\label{particular transformation}

 Let us now study a particular case of the
transformation (\ref{transf inversa}). Taking $b_1=1$ and $b_2=\frac{1}{\gamma}$ the transformation is invertible for
$\gamma^2\neq \sigma$. In this case, the action (\ref{cmpr1}) takes the form
\begin{gather}
S[B,A,\phi,\mu] =\int_{\mathcal{M}^4} \bigg[ \left(B^{IJ}+\frac1{\gamma}\,{^*B}^{IJ}\right) \wedge F_{IJ} [A] -
\frac12 \phi_{IJKL} B^{IJ} \wedge B^{KL} \nonumber \\
\hphantom{S[B,A,\phi,\mu] =\int_{\mathcal{M}^4} \bigg[ }{} -   \mu  \left ( A_1  \phi_{IJ}{}^{IJ} + A_2
\phi_{IJKL}  \varepsilon^{IJKL} \right ) \bigg],\label{cmpr hacia eprl}
\end{gather}
where now
\begin{gather}
A_1=\frac{\gamma^2}{(\gamma^2-\sigma)^2}\left[a_1 \left(\gamma^2+\sigma\right)-4\sigma a_2\gamma\right], \qquad
\label{A's para eprl1}  A_2=\frac{\gamma^2}{(\gamma^2-\sigma)^2}\left[a_2
\left(\gamma^2+\sigma\right)-a_1\gamma\right].
\end{gather}
It is important to notice that, at this stage, the action principle (\ref{cmpr hacia eprl}) is completely equivalent to
action (\ref{cmpr1}) because the coef\/f\/icients of the transformation
satisfy the condition (\ref{condition}).

As pointed out in the previous remark (b), it is possible to obtain only one of the invariants in the action (\ref{cmpr
hacia eprl}) by imposing, additionally, either $A_1=0$ or $A_2=0$.

\subsubsection[Case $A_1=0$: action with the invariant $\phi_{IJKL}  \varepsilon^{IJKL}$ only]{Case $\boldsymbol{A_1=0}$: action with the invariant $\boldsymbol{\phi_{IJKL}  \varepsilon^{IJKL}}$ only}
\label{particular transformation epr}

 In order to eliminate the term with $\phi_{IJ}{}^{IJ}$ in the action (\ref{cmpr
hacia eprl}), $A_1$ must vanish; this is only possible if $a_2$ and $a_1$ satisfy the condition
\begin{gather}\label{condition eprl}
\frac{a_2}{a_1}=\frac{\gamma^2+\sigma}{4\gamma\sigma}.
\end{gather}
Using (\ref{condition eprl}), the CMPR action principle (\ref{cmpr1}) takes the form
\begin{gather}
S[Q,A,\psi,\mu] =\int_{\mathcal{M}^4} \bigg[ Q^{IJ} \wedge F_{IJ} [A] - \frac12 \psi_{IJKL} Q^{IJ} \wedge Q^{KL}
  \nonumber \\
  \hphantom{S[Q,A,\psi,\mu] =\int_{\mathcal{M}^4} \bigg[}{}
  -    \mu a_1  \left( \psi_{IJ}{}^{IJ} + \frac{\gamma^2+\sigma}{4\gamma\sigma} \psi_{IJKL}
\varepsilon^{IJKL} \right) \bigg],\label{key}
\end{gather}
whereas the transformed action principle (\ref{cmpr hacia eprl}) becomes
\begin{gather}
S[B,A,\phi,\mu] =\int_{\mathcal{M}^4} \bigg[ \left ( B^{IJ} + \frac{1}{\gamma}\, {^{\ast} B}^{IJ} \right ) \wedge
F_{IJ} [A] - \frac12 \phi_{IJKL} B^{IJ} \wedge B^{KL}   \nonumber \\
\hphantom{S[B,A,\phi,\mu] =\int_{\mathcal{M}^4} \bigg[ }{}
-
 \mu \frac{ a_1 \sigma  \gamma}{4}   \phi_{IJKL}  \varepsilon^{IJKL} \bigg].\label{eprl1}
\end{gather}
This is the form of the action principle used in~\cite{epr} (see also~\cite{oriti}).

\pagebreak

\noindent
{\it Remarks:}
\begin{enumerate}\itemsep=0pt
\setcounter{enumi}{2}
\item In the Lorentzian case $\sigma=-1$, the particular values
    $\gamma=\pm1$ imply $\gamma^2+\sigma=0$ and thus the Lorentz invariant $\phi_{IJKL}  \varepsilon^{IJKL}$ is {\it not} present
    in the action (\ref{key}), which then reduces to the form given in (\ref{traza}). This means that the action
    (\ref{eprl1}) can be written as the action studied in~\cite{capo} by taking $\gamma=\pm 1$ in the Lorentzian case.
    For any other arbitrary real value of the Immirzi parameter the two invariants $\psi_{IJ}{}^{IJ}$ and $\psi_{IJKL}
    \varepsilon^{IJKL}$ are present in the action~(\ref{key})~\cite{cqgl2001}.

\item In the Euclidean case, $\sigma=1$, for {\it real} values of $\gamma$
    it follows that $\gamma^2+\sigma \neq0$ and therefore the two Lorentz
    invariants are always present in the action (\ref{key}). This means
    that the {\it two} invariants $\psi_{IJ}{}^{IJ}$ and $\psi_{IJKL}
    \varepsilon^{IJKL}$ must be involved in order to include arbitrary
    real values of $\gamma$, as it was recognized in~\cite{cqgl2001}.
    Nevertheless, it is important to notice that $\gamma^2+\sigma$ can
    vanish if complex values of $\gamma$ are allowed, $\gamma=\pm i$. For
    these values the invariant $\psi_{IJKL} \varepsilon^{IJKL}$ is
    missing in the action~(\ref{key}), which becomes also the one given in
    (\ref{traza}).
\end{enumerate}

Continuing with the analysis, the expression for the $B$'s can be directly obtained from action principle
(\ref{eprl1}). However, it can be alternatively obtained from the $Q$'s given in (\ref{sol}) and (\ref{quotien}) (and
supplemented with (\ref{condition eprl})) and from the use of the inverse transformation (\ref{transformacion}) with
$b_1=1$ and $b_2=\frac{1}{\gamma}$. We are going to follow this last approach. Therefore, from the equality of equations
(\ref{quotien}) and (\ref{condition eprl}) it follows that in order for the $Q$'s in (\ref{sol}) to be solutions for
the action principle (\ref{key}), $\alpha/\beta$ must satisfy the quadratic equation
\[
\left (\frac{\alpha}{\beta} \right )^2 - \left ( \frac{1}{\gamma} + \sigma \gamma \right ) \frac{\alpha}{\beta} +
\sigma =0,
\]
whose solutions are
\begin{gather}\label{roots}
(i) \quad \alpha/\beta=  \sigma \gamma, \qquad \mbox{and} \qquad (ii) \quad \alpha/\beta = \frac{1}{\gamma}.
\end{gather}
The f\/irst root was explicitly mentioned in~\cite{cqgl2001}, but the second one was not recognized there as a
possibility to include the Immirzi parameter.

Inserting the two roots given in (\ref{roots}) into (\ref{sol}), we get the corresponding expression for the $Q$'s
\begin{gather}
 (i) \quad Q^{IJ}= \alpha \left[\, {^{\ast}} \left( e^I \wedge e^J \right ) + \frac{\sigma}{\gamma} e^I \wedge e^J
\right],
\qquad \label{Q's de eprl}
 (ii) \quad Q^{IJ}= \alpha \left [\, {^{\ast}} \left ( e^I \wedge e^J \right ) + \gamma e^I \wedge e^J \right ],
\end{gather}
and by plugging them into (\ref{transformacion}) with the restrictions $b_1=1$ and $b_2=\frac{1}{\gamma}$ we get the
corresponding expressions for the $B$'s
\begin{gather}
(i) \quad B^{IJ}= \alpha\, {^{\ast}} \left ( e^I \wedge e^J \right ),
\qquad
(ii) \quad B^{IJ}=  \alpha \gamma   e^I \wedge e^J, \label{clave}
\end{gather}
which are precisely the ones given in (\ref{pietriEXTRA}).
 Furthermore, by comparing (\ref{pietriEXTRA}) and (\ref{clave}),
 we conclude that $\kappa_1 = \alpha$ for $(i)$ (and thus $\kappa_1=\sigma\gamma\beta$)
 whereas $\kappa_2 = \alpha \gamma $ for $(ii)$ (and thus $\kappa_2=\beta$).

 By plugging (\ref{clave}) into (\ref{eprl1}) or, equivalently, by plugging (\ref{Q's de eprl}) into
(\ref{key}), we get
\begin{gather*}
(i) \quad S_1[e,A]=\kappa_1 \int_{{\mathcal M}^4}\left[{^{\ast}} \left( e^I \wedge e^J \right) +\frac{\sigma}{\gamma}
e^I \wedge e^J \right]\wedge F_{IJ}[A], \\
(ii) \quad  S_2[e,A]=\frac{\kappa_2}{\gamma} \int_{{\mathcal M}^4}\left[\,{^{\ast}} \left(e^I \wedge e^J\right) +
\gamma  e^I \wedge e^J  \right]\wedge F_{IJ}[A],
\end{gather*}
which is exactly the same result that we had obtained if we had directly solved the constraint on the $B$'s that comes
from~(\ref{eprl1})~\cite{epr}. It is common to take $\kappa_1=\pm 1$ and $\kappa_2 = \pm 1$. Nevertheless, it must be
stressed that these values do not come out from the sole handling of the equations of motion.

\subsubsection[Case $A_2=0$: action with the invariant $\phi_{IJ}{}^{IJ}$ only]{Case $\boldsymbol{A_2=0}$: action with the invariant $\boldsymbol{\phi_{IJ}{}^{IJ}}$ only}

 In order to obtain the action (\ref{eprl1}) from (\ref{cmpr hacia eprl}), a particular function for the ratio $a_2/a_1$
has been taken such that $A_1$ in equation~(\ref{A's para eprl1}) vanishes once the transformation is performed. So, it is
natural to ask what happens if, instead of $A_1$, it is the coef\/f\/icient $A_2$ which is forced to vanish in such a way
that the invariant $\phi_{IJKL}\varepsilon^{IJKL}$ is not present in the action (\ref{cmpr hacia eprl}). From (\ref{A's
para eprl1}) this condition is equivalent to
\begin{gather}\label{condition S1}
\frac{a_1}{a_2}=\frac{\gamma^2+\sigma}{\gamma}.
\end{gather}
This means that starting from the CMPR action principle in the form
\begin{gather}
S[Q,A,\psi,\mu] =\int_{\mathcal{M}^4} \bigg[ Q^{IJ} \wedge F_{IJ} [A] - \frac12 \psi_{IJKL} Q^{IJ} \wedge Q^{KL}
  \nonumber \\
 \phantom{S[Q,A,\psi,\mu] =\int_{\mathcal{M}^4} \bigg[}{} -   \mu a_2  \left (\frac{\gamma^2+\sigma}{\gamma} \psi_{IJ}{}^{IJ} + \psi_{IJKL}
\varepsilon^{IJKL} \right ) \bigg],\label{key2}
\end{gather}
and using the transformation (\ref{transformacion}) with $b_1=1$ and $b_2=1/\gamma$, this action principle acquires the
form
\begin{gather}
S[B,A,\phi,\mu] =\int_{\mathcal{M}^4} \!\left [ \left ( B^{IJ} + \frac{1}{\gamma} \,{^{\ast} B}^{IJ} \right )\! \wedge
F_{IJ} [A] - \frac12 \phi_{IJKL} B^{IJ} \wedge B^{KL}
 - \mu a_2 \gamma  \phi_{IJ}{}^{IJ} \right ].\!\!\!\label{mmv}
\end{gather}
{\it Remarks:}
\begin{enumerate}\itemsep=0pt
\setcounter{enumi}{4}
\item In the case $\gamma^2+\sigma\neq0$ it follows from (\ref{key2}) that
    the two Lorentz invariants are present in the action and therefore the
    expression for the two-forms $Q$'s in terms of the tetrad f\/ield, and
    thus the form of the $B$'s in~(\ref{mmv}), can be obtained from~(\ref{sol}) following the same procedure carried out in the previous
    section. In this case the value of $a_2/a_1$ comes from~(\ref{condition S1}) and the f\/inal form of the actions~(\ref{key2})
    and~(\ref{mmv}) in terms of the tetrad f\/ield and the connection are
    \begin{gather*}
        (i)\quad S_1 [e,A]= \alpha \int_{\mathcal{M}^4} \left[ \,{^{\ast}} \left(e^{I}\wedge e^{J} \right)
        +\sqrt{-\sigma} \frac{\gamma-\sqrt{-\sigma}}{\gamma+\sqrt{-\sigma}} e^I\wedge e^J  \right] \wedge F_{IJ}
        [A],\\
        (ii)\quad S_2 [e,A]= \alpha \int_{\mathcal{M}^4} \left[ \,{^{\ast}} \left(e^{I}\wedge e^{J} \right)
        -\sqrt{-\sigma} \frac{\gamma+\sqrt{-\sigma}}{\gamma-\sqrt{-\sigma}}  e^I\wedge e^J \right] \wedge F_{IJ}
        [A],
    \end{gather*}
    corresponding to the two solutions for the $B$'s.
\item In the case $\gamma^2 + \sigma =0$ the Lorentz invariant $\psi_{IJ}{}^{IJ}$
    is missing in the action (\ref{key2}) and it becomes the one given in
    (\ref{epsilon}). In that analysis the action principles (\ref{epsilonENes}) were obtained by plugging in
    (\ref{epsilon}) the solution for the $Q$'s. In a similar way, the action principle (\ref{mmv}) restricted to $\gamma =\pm \sqrt{-\sigma}$
    can be written in terms of the tetrad f\/ield by solving for the $B$'s and plugging the solutions into (\ref{mmv}). The resulting action
    principles obtained following one or the other procedure
    are dif\/ferent by a global factor.
\end{enumerate}

\section{Application of the transformation to the coupling\\ of the cosmological constant}\label{SA}

The coupling of the cosmological constant to the CMPR action principle was done in~\cite{prd2010} (see also~\cite{simone}). However, taking into account the previous analysis, it is natural to wonder how the coupling looks like
in the transformed CMPR action principles discussed in Section~\ref{trasfCMPR}. In particular we are interested in the
action principle analyzed in Section~\ref{particular transformation epr} and used in~\cite{epr}. To do this
task, the starting point is the CMPR action principle coupled with the cosmological constant given in~\cite{prd2010}. This action principle can be transformed to one with {\it two} BF terms by applying the general
transformation encoded in equations~(\ref{transf inversa}) and~(\ref{psi en phi}). This will give us an equivalent action
principle for gravity with cosmological constant. Nevertheless, as we are interested in the coupling of the
cosmological constant to gravity in an action principle of the form given in (\ref{eprl1}), i.e.\ with the linear
combination $B^{IJ} + \frac{1}{\gamma}\,^{\ast}B^{IJ}$ and using only the Lorentz invariant
$\phi_{IJKL}\varepsilon^{IJKL}$, it is necessary to apply the particular transformation studied in Section~\ref{trasfCMPR} (def\/ined by $b_1=1$ and $b_2=\frac{1}{\gamma}$), and to restrict the constants included in the CMPR
action principle to those which lead us to (\ref{eprl1}), i.e.\ to impose on $a_1$ and $a_2$ the condition~(\ref{condition eprl}).

The action principle introduced in~\cite{prd2010} is given by
\begin{gather}
 S [Q,A,\psi,\mu] =\int_{\mathcal{M}^4} \Big [ Q^{IJ} \wedge F_{IJ} [A] - \frac12 \psi_{IJKL} Q^{IJ} \wedge Q^{KL}
 \nonumber\\
 \phantom{S [Q,A,\psi,\mu] =}{}
 - \mu  \left ( a_1  \psi_{IJ}{}^{IJ} + a_2  \psi_{IJKL} \varepsilon^{IJKL} - \lambda\right )
+ l_1 Q_{IJ} \wedge Q^{IJ} + l_2 Q_{IJ} \wedge ^{\ast}Q^{IJ} \Big],\label{cmpr+cc}
\end{gather}
with $\lambda=a_1(4!l_2\sigma \frac{a_2}{a_1}+12 l_1+\frac{\Lambda}{\beta})$. It is important to mention that this
expression for $\lambda$ comes from the fact that $\Lambda$ is identif\/ied with the cosmological constant in the tetrad
formalism (see~\cite{prd2010} for the details).

By applying the transformation given in equation (\ref{transf inversa}), the action principle (\ref{cmpr+cc}) takes the form
\begin{gather}
 S[B,A,\phi,\mu] =\int_{\mathcal{M}^4} \Big [ \left(b_1 B^{IJ}+ b_2 \, {^*}B^{IJ}\right) \wedge F_{IJ} [A] - \frac12
\phi_{IJKL} B^{IJ} \wedge B^{KL}  \label{accion transformada+CC} \\
\phantom{S[B,A,\phi,\mu] =}{}
  - \mu  \left ( A_1\, \phi_{IJ}{}^{IJ} + A_2  \phi_{IJKL}
\varepsilon^{IJKL}-\lambda \right ) + K_1 B_{IJ} \wedge B^{IJ} + K_2 B_{IJ} \wedge ^{\ast}B^{IJ} \Big],\nonumber
\end{gather}
where the constants $A_1$ and $A_2$ are given by equation~(\ref{config1}) and
\begin{gather}
K_1 = l_1 \big(b_1^2+\sigma b_2^2\big) + 2 l_2 b_1 b_2 \sigma, \qquad
K_2 = l_2 \big(b_1^2+\sigma b_2^2\big) + 2 l_1 b_1 b_2. \label{config2}
\end{gather}
Note that the last two terms of equation~(\ref{accion transformada+CC}) come from the analogous terms that appear in equation~(\ref{cmpr+cc}). These terms could still be removed by doing a redef\/inition of the Lagrange multiplier
$\phi_{IJKL}\rightarrow\varphi_{IJKL}$ but this would imply a transformation $\psi_{IJKL}\rightarrow\varphi_{IJKL}$
dif\/ferent from the one given in~(\ref{phi en psi}). Such a transformation would be an example of how we can choose the
transformation depending on the action principles we want to relate. Nevertheless, in this section we are only
interested in an application of the transformation introduced in Section~\ref{trasfCMPR}, therefore we will continue
with the original transformation~(\ref{phi en psi}).

Notice that $\lambda$ can be written in terms of $K_1$ and $K_2$ as
\begin{gather*}
\lambda= 12 A_1 K_1+4!\sigma A_2 K_2 + \frac{\Lambda a_1}{\beta}. 
\end{gather*}
Let us now take the particular transformation $b_1=1$ and $b_2=\frac{1}{\gamma}$ used in Section~\ref{particular
transformation epr}.
In this case, the action~(\ref{accion transformada+CC}) takes on the form
\begin{gather}
 S[B,A,\phi,\mu] =\int_{\mathcal{M}^4} \bigg[ \left(B^{IJ}+\frac1{\gamma}\,{^*B}^{IJ}\right) \wedge F_{IJ} [A] -
\frac12 \phi_{IJKL} B^{IJ} \wedge B^{KL}\label{cmpr+cc hacia eprl+cc}
  \\
  \phantom{S[B,A,\phi,\mu] =}{}
 - \mu  \left( A_1  \phi_{IJ}{}^{IJ} + A_2  \phi_{IJKL} \varepsilon^{IJKL}-\lambda \right) + K_1 B_{IJ} \wedge B^{IJ} + K_2 B_{IJ}
\wedge ^{\ast}B^{IJ} \bigg],\nonumber
\end{gather}
where now $A_1$ and $A_2$ are given by (\ref{A's para eprl1}), and (\ref{config2}) reduces to
\begin{gather*}
K_1 = l_1\frac{(\gamma^2+\sigma)^2}{\gamma^2}+2 l_2\frac{\sigma}{\gamma}, \qquad
K_2 = l_2\frac{(\gamma^2+\sigma)^2}{\gamma^2}+2 l_1\frac{1}{\gamma}. 
\end{gather*}
It is important to note that, at this stage, the action principle~(\ref{cmpr+cc hacia eprl+cc}) is completely
equivalent to action (\ref{cmpr+cc}) because the coef\/f\/icients of the transformation $b_1$ and $b_2$ satisfy the
condition~(\ref{condition}).

In order to eliminate the term with $\phi_{IJ}{}^{IJ}$ in the action (\ref{cmpr+cc hacia eprl+cc}),
$a_1$ and $a_2$ must satisfy the condition (\ref{condition eprl}) which f\/ixes the constants $A_1$ and $A_2$ to
\begin{gather*}
A_1=0, \qquad \mbox{and} \qquad A_2=\frac{a_1\gamma\sigma}{4},
\end{gather*}
while the constants $K_1$ and $K_2$ do not get modif\/ied. Using the values for $A_1$ and $A_2$ the action principle
(\ref{cmpr+cc hacia eprl+cc}) takes the form
\begin{gather}\nonumber
S[B,A,\phi,\mu]  = \int_{\mathcal{M}^4} \bigg[ \left(B^{IJ}+\frac1{\gamma}\,{^*B}^{IJ}\right) \wedge F_{IJ} [A] -
\frac12 \phi_{IJKL} B^{IJ} \wedge B^{KL}   \\
\phantom{S[B,A,\phi,\mu]  =}{} - \mu  \left (\frac{a_1 \gamma\sigma}{4}  \phi_{IJKL}
\varepsilon^{IJKL}-\lambda \right ) + K_1 B_{IJ} \wedge B^{IJ} + K_2 B_{IJ} \wedge ^{\ast}B^{IJ} \bigg],\label{eprl+cc}
\end{gather}
and $\lambda$ becomes
\begin{gather}\label{H en K's}
\lambda=a_1\left[ 3!\gamma K_2+\frac{\Lambda}{\beta}\right].
\end{gather}
Because of the restriction (\ref{condition eprl}), the action principle (\ref{eprl+cc}) is a particular case of
(\ref{cmpr+cc hacia eprl+cc}) and thus a~particular case of (\ref{cmpr+cc}).

In order to write the action principle (\ref{eprl+cc}) in terms of the tetrad and a connection we can follow two
approaches. In the f\/irst, we simply use the form for the $B$'s obtained in (\ref{clave}) and the value of $\lambda$
given in (\ref{H en K's}) and insert them into (\ref{eprl+cc}) to get
\begin{gather*}
 (i)\quad S[e,A]=\alpha \int_{{\mathcal M}^4}\left[\left({^{\ast}} \left( e^I \wedge e^J \right) +\frac{\sigma}{\gamma} e^I \wedge e^J
\right)\wedge F_{IJ}[A]-\frac{\Lambda}{12}\varepsilon_{IJKL} e^I\wedge e^J\wedge e^K\wedge e^L\right],
\nonumber \\ 
 (ii)\quad S[e,A]=\alpha \int_{{\mathcal M}^4}\left[\left( \,{^{\ast}} \left(e^I \wedge e^J\right) + \gamma
 e^I \wedge e^J  \right)\wedge F_{IJ}[A] -\frac{\Lambda}{12}\varepsilon_{IJKL} e^I\wedge e^J\wedge e^K\wedge e^L\right],
\end{gather*}
which is the Holst action principle with cosmological constant. Note that each solution involves a dif\/ferent Immirzi
parameter.

In the second approach, we start from the action principle (\ref{eprl+cc}) where now $\lambda$ is not f\/ixed and we
simply solve the constraint on the $B$'s that comes from (\ref{eprl+cc}). The relation of $\lambda$ with the cosmological
constant will be obtained at the end of the procedure. The equation of motion that comes from the variation of~(\ref{eprl+cc}) with respect to the f\/ield $\phi_{IJKL}$ implies
\begin{gather*}
  \frac12 B_{IJ}\wedge B_{KL}= -\frac{\mu a_1 \gamma\sigma}{4}\varepsilon_{IJKL}, 
\end{gather*}
and therefore
\begin{gather}
B_{IJ}\wedge B^{IJ}=0, 
\qquad
B_{IJ}\wedge ^{\ast}B^{IJ}=-3! \mu a_1 \gamma. \label{eq motion 2}
\end{gather}
From which we obtain the two-forms $B$'s given by
\begin{gather}\label{clave2}
(i)\quad B^{IJ}=\kappa_1\, ^{\ast}\left(e^I \wedge e^J \right) \qquad \mbox{and}  \qquad (ii) \quad B^{IJ}=\kappa_2 \left(e^I
\wedge e^J\right).
\end{gather}
Using equations (\ref{eq motion 2}) and (\ref{clave2})
we obtain
\begin{gather*}\nonumber
 (i) \quad S[e,A]=\kappa_1\int_{\mathcal{M}^4} \bigg[\left( ^{\ast}\left(e^I\wedge e^J\right) +\frac{\sigma}{\gamma} e^I \wedge
e^J \right) \wedge F_{IJ} [A]   \\ 
\phantom{(i) \quad S[e,A]=}{}  + \frac{\kappa_1 \sigma}{2}
 \left( K_2 -\frac{1}{3!\gamma}\frac{\lambda}{a_1}\right) \varepsilon_{IJKL} e^I\wedge e^J \wedge e^K \wedge e^L\bigg],\\
 (ii) \quad S[e,A]=\frac{\kappa_2}{\gamma}\int_{\mathcal{M}^4} \bigg[\left( ^{\ast}\left(e^I \wedge e^J\right) +
\gamma\,e^I\wedge e^J \right) \wedge F_{IJ} [A]   \nonumber \\
\phantom{ (ii) \quad S[e,A]=}{}  + \frac{\kappa_2\gamma}{2}
 \left( K_2 - \frac{1}{3!\gamma}\frac{\lambda}{a_1}\right) \varepsilon_{IJKL} e^I\wedge e^J \wedge e^K \wedge e^L\bigg]. 
\end{gather*}
These actions have the form of the Holst's action principle with cosmological constant $\Lambda$ given in each case by
\begin{gather*}
 (i)\quad \frac{\Lambda}{12}=- \frac{\kappa_1 \sigma}{2} \left( K_2 - \frac{1}{3!\gamma}\frac{\lambda}{a_1}\right),\qquad
 (ii)\quad \frac{\Lambda}{12}=- \frac{\kappa_2 \gamma}{2} \left( K_2 - \frac{1}{3!\gamma}\frac{\lambda}{a_1}\right).
\end{gather*}
They f\/ix the relationship among the constants $\lambda$, $\gamma$, and $a_1$ that appear in the action princip\-le~(\ref{eprl+cc}), and the coef\/f\/icients of the solutions given in (\ref{clave2}). It is easy to see that, for each case,
$\lambda$~has the form
\begin{gather*}
 (i)\quad \lambda=a_1\left[3! \gamma K_2 +\frac{\Lambda \sigma \gamma}{\kappa_1}\right], \qquad
 (ii)\quad \lambda=a_1\left[3!\gamma K_2 +\frac{\Lambda }{\kappa_2}\right],
\end{gather*}
respectively.
Note that in order for these values of $\lambda$ match the value of $\lambda$ given in (\ref{H en K's}) (i.e.\ the two
approaches give the same result), it is required that $\kappa_1=\sigma\gamma\beta$ and $\kappa_2= \alpha \gamma$,
respectively. These are the same values obtained for $\kappa_1$ and $\kappa_2$ in the analysis after equation~(\ref{clave}).

\section{Conclusions}

It has been shown that by performing an invertible transformation of the f\/ields of the BF formulation for general
relativity given by Capovilla, Montesinos, Prieto, and Rojas \cite{cqgl2001}, it is possible to obtain the action
principle (\ref{accion transformada}) which includes the two BF terms $B^{IJ}\wedge F_{IJ}$ and $^{\ast}B^{IJ}\wedge
F_{IJ} $ and still involves the two Lorentz invariants $\phi_{IJKL} \varepsilon^{IJKL}$ and $\phi_{IJ}{}^{IJ}$
generically. One of the results of the analysis is to clearly show the relationship of the two parameters $a_1$ and
$a_2$ of the CMPR action principle and of the two parameters $b_1$ and $b_2$ involved in the transformation with the
Immirzi parameter.

From the analysis is clear that the freedom in the choice of the parameters $a_1$, $a_2$,~$b_1$, and~$b_2$ can be used
to handle the two Lorentz invariants $\phi_{IJKL} \varepsilon^{IJKL}$ and $\phi_{IJ}{}^{IJ}$ that appear in the
transformed action principle. In particular, a suitable combination of these parameters can result in that one of these
invariants is missing in the transformed action principle as it is shown in the Section~\ref{particular
transformation}. As an application of this fact, the action used in~\cite{epr} is obtained in Section~\ref{particular transformation epr}.

Finally, and as another application of the transformation discussed in this paper, the coupling of the cosmological
constant to the action principle used in~\cite{epr} is obtained from the coupling of the cosmological constant to
the CMPR action principle studied in~\cite{prd2010}.

 \vspace{-2mm}

\subsection*{Acknowledgements}
This work was partially supported by Conacyt, grant number 56159-F.

 \vspace{-2mm}

\pdfbookmark[1]{References}{ref}
\LastPageEnding

\end{document}